\documentclass[a4paper,fleqn,usenatbib]{mnras}

\usepackage{txfonts}

\usepackage[T1]{fontenc}
\usepackage{ae,aecompl}
\usepackage{multirow}

\usepackage{graphicx}	
\usepackage{amssymb}	

\usepackage{tikz}
\usetikzlibrary{arrows}

\title[MNPs and X-class flares]{Properties of magnetic null points associated with X-class flares during solar cycle 24}
\author[Edgar and R{\'e}gnier]{
R. L. Edgar,$^{1}$ and
S. R{\'e}gnier$^{1}$\thanks{E-mail: stephane.regnier@northumbria.ac.uk}\\
$^{1}$Department of Mathematics, Physics, and Electrical Engineering, Northumbria University, Newcastle upon Tyne,
NE1 8ST, UK \\
\date{Accepted XXX. Received YYY; in original form ZZZ}
}

\begin{document}
\label{firstpage}
\pagerange{\pageref{firstpage}--\pageref{lastpage}}
\maketitle

\begin{abstract}
Since the launch of the Solar Dynamics Observatory (SDO) in 2010 and throughout the solar cycle 24, the Sun has produced few tens of X-class ﬂares, which are the most energetic solar events. Those flares are produced in regions where the magnetic flux/energy is large and the magnetic configurations are complex. To provide more insights into the flaring process, we investigate the properties of magnetic null points (MNPs) and their correlation with the energy release sites. During solar cycle 24, we identify 17 X-class flares satisfying selection criteria. From {\em SDO}/HMI magnetograms, we perform potential extrapolations around the peak time of the ﬂare to access the 3D coronal magnetic ﬁeld and thus investigate the existence of coronal MNPs. We then correlate the flaring sites with the existing MNPs using {\em SDO}/AIA 171~$\AA$ EUV observations, and deduce their properties (sign, spine, fan). Six active regions out of 10 possess at least one MNP  which is stable and with large magnetic field gradients: this implies that 35\% of X-class ﬂares are associated with a MNP; of which 87.5\% of MNPs are of positive type. The MNPs associated with the flare sites are predominantly located at a height between 0.5 and 2 Mm, and with a vertical/radial spine field line. We also ﬁnd a slight correlation between the MNPs not associated with a ﬂare and negative-type MNPs (55\%) within the active region. Regarding the physics of flares, the association between the enhanced intensity at the flaring site and a MNP represents about a third of the possible scenarios for triggering X-class flares.
\end{abstract}

\begin{keywords}
Sun: magnetic fields -- Sun: corona -- Sun: flares
\end{keywords}



\section{Introduction}\label{sec:intro}

\begin{table*}
	\centering
	\caption{Properties of the 17 X-class flares selected}
	\label{tab:events}
	\begin{tabular}{cccccccc} 
		\hline
		\hline
		Date & Start Time & Peak Time & End Time & Class & Source & AR location & Event Number \\
		& (UT) & (UT) & (UT) & & NOAA \# & & \citet{2017ApJ...834...56T} \\
		\hline
		2011 Feb 15 & 01:44 & 01:56 & 02:06 & X2.2 & AR 11158 & S21W28 & 2 \\
		2011 Mar 9 & 23:13 & 23:23 & 23:29 & X1.5 & AR 11166 & N11W15 & 3 \\
		2011 Sep 6 & 22:12 & 22:20 & 22:24 & X2.1 & AR 11283 & N14W18 & 8 \\
		2011 Sep 7 & 22:32 & 22:38 & 22:44 & X1.8 & AR 11283 & N14W32 & 9 \\
		2012 Mar 7 & 00:02 & 00:24 & 00:40 & X5.4 & AR 11429 & N17E15 & 11 \\
		2012 Mar 7 & 01:05 & 01:14 & 01:23 & X1.3 & AR 11429 & N17E15 & 12 \\
		2012 Jul 12 & 15:37 & 16:49 & 17:30 & X1.4 & AR 11520 & S17E08 & 19 \\
		2013 Nov 5 & 22:07 & 22:12 & 22:15 & X3.3 & AR 11890 & S09E36 & 24 \\
		2013 Nov 8 & 04:20 & 04:26 & 04:29 & X1.1 & AR 11890 & S11W03 & 25 \\
		2013 Nov 10 & 05:08 & 05:14 & 05:18 & X1.1 & AR 11890 & S11W28 & 26 \\
		2014 Mar 29 & 17:35 & 17:48 & 17:54 & X1.0 & AR 12017 & N10W32 & 31 \\
		2014 Sep 10 & 17:21 & 17:45 & 18:20 & X1.6 & AR 12158 & N15E02 & 33 \\
		2014 Oct 22 & 14:02 & 14:28 & 14:50 & X1.6 & AR 12192 & S14 E05 & 36 \\
		2014 Oct 24 & 21:07 & 21:41 & 22:13 & X3.1 & AR 12192 & S14W20 & 37 \\
		2014 Oct 25 & 16:55 & 17:08 & 18:11 & X1.0 & AR 12192 & S12W35 & 38 \\
		2014 Oct 26 & 10:04 & 10:56 & 11:18 & X2.0 & AR 12192 & S12W46 & 39 \\
		2014 Dec 20 & 00:11 & 00:28 & 00:55 & X1.8 & AR 12242 & S18W42 & 45 \\
		\hline
	\end{tabular}
\end{table*}

The solar dynamo is amplifying and storing magnetic energy at the interface between the radiative and convection zones. The stored energy is then transported towards the top of the convection zone and then penetrates the photosphere and the upper layers of solar atmosphere. Finally, the excess energy is re-distributed locally and/or at large scale within the heliosphere through flares and Coronal Mass Ejections (CMEs). Observations provide a wealth of information on the structure of the solar atmosphere to help understand the processes responsible for releasing the excess magnetic energy. In particular, it is important to understand the complexity of the magnetic field configurations that often exhibits a magnetic topology (e.g. magnetic null points, separatrices), an intricated connectivity, and a clear localised distribution of magnetic helicity. However, it is still unclear if there is a direct link between the magnetic topology characterised by a discontinuity in the mapping of the field and the sites of magnetic energy released in the solar corona. We here investigate the correlation between the existence of magnetic null points and the coronal brightenings defining the location of the flaring activity.   

Several works have investigated the relationship between the complexity of the magnetic field and the activity of active regions (ARs). \citet{2017ApJ...834...56T} have studied a statistically significant number of flares (51 flares above M5.0 class) and their relationship with the complex distribution of the magnetic field in the photosphere. The authors concluded that X-class flares do not always require a complex photospheric field distribution (i.e. a $\delta$-sunspot) or strong magnetic gradients at the polarity inversion line. This suggest that the photospheric magnetic field alone does not reveal the full complexity of the magnetic field, and that it is crucial to access to the 3D magnetic field configuration within the solar atmosphere. Currently, only the coronal magnetic field extrapolation methods are providing such information in the limit of their physics assumptions.

\citet{2007ApJ...662.1293U} studied the relationship between flares and CMEs with an emphasis on the magnetic configuration of the CME's source regions. The authors proceed to the potential field extrapolations of {\em SOHO}/MDI magnetograms (SOlar and Heliospheric Observatory/Michelson Doppler Imager, \citeauthor{1995SoPh..162..129S} \citeyear{1995SoPh..162..129S}) and then locate possible MNPs in the 3D magnetic configurations. Potential magnetic field extrapolations are performed at a single time before, during or after the flare. The events cover a period from 1998 to 2002 during the rising phase and the first peak of activity of solar cycle 23. For the events selected, CME sources, i. e. active regions observed in the low solar atmosphere, are producing eruptions from C-class to X-class flares. Amongst the 12 X-class flares, 7 flares possess a MNP associated with the erupting active region (58\%), and 5 flares are without a MNP (42\%). As the authors focused on the trigger mechanisms of CMEs (and not of flares), the spatial correlation between the location of the flare and the location of the MNPs present in the magnetic configuration is not investigated.  

In \citet{2007ApJ...662.1293U}, the focus was on relating the eruptive flares with the breakout model \citep{1999ApJ...510..485A} as a universal mechanism for flares. The authors concluded that there is not a unique mechanism that trigger flares and CMEs.  

The nature of MNPs in the corona has also been studied at large-scales. \citet{2018PhDT.......166W} studied the topology of the coronal magnetic fields using a Potential Field Source Surface model\citep{1969SoPh....9..131A,1969SoPh....6..442S}, thus finding the topological skeletons during the solar cycle 24. The author found a large number of magnetic null points or magnetic null lines in the corona, even if the MNP's density is low for the volume considered. These results are consistent with the different studies for the quiet-Sun topology from the photosphere to the corona \citep[see e. g.][]{2008A&A...484L..47R,2009SoPh..254...51L}.  

In this paper, the selection and analysis of the events follow a similar approach to \citet{2007ApJ...662.1293U}. We restrict our study to X-class flares observed during the solar cycle 24 by the Solar Dynamics Observatory ({\em SDO}, \citeauthor{2012SoPh..275....3P} \citeyear{2012SoPh..275....3P}). The potential field extrapolations are using {\em SDO}/HMI \citep{2012SoPh..275..207S} photospheric magnetograms as bottom boundary conditions, and {\em SDO}/AIA \citep{2012SoPh..275...17L} images to confirm the location of the flares (see Section~\ref{sec:data}). In Section~\ref{sec:res}, we provide a statistical study of the correlation between the MNP's properties and the flares, before drawing the conclusions in Section~\ref{sec:concl}.

\begin{figure*}
\centering
\includegraphics[width=.9\linewidth]{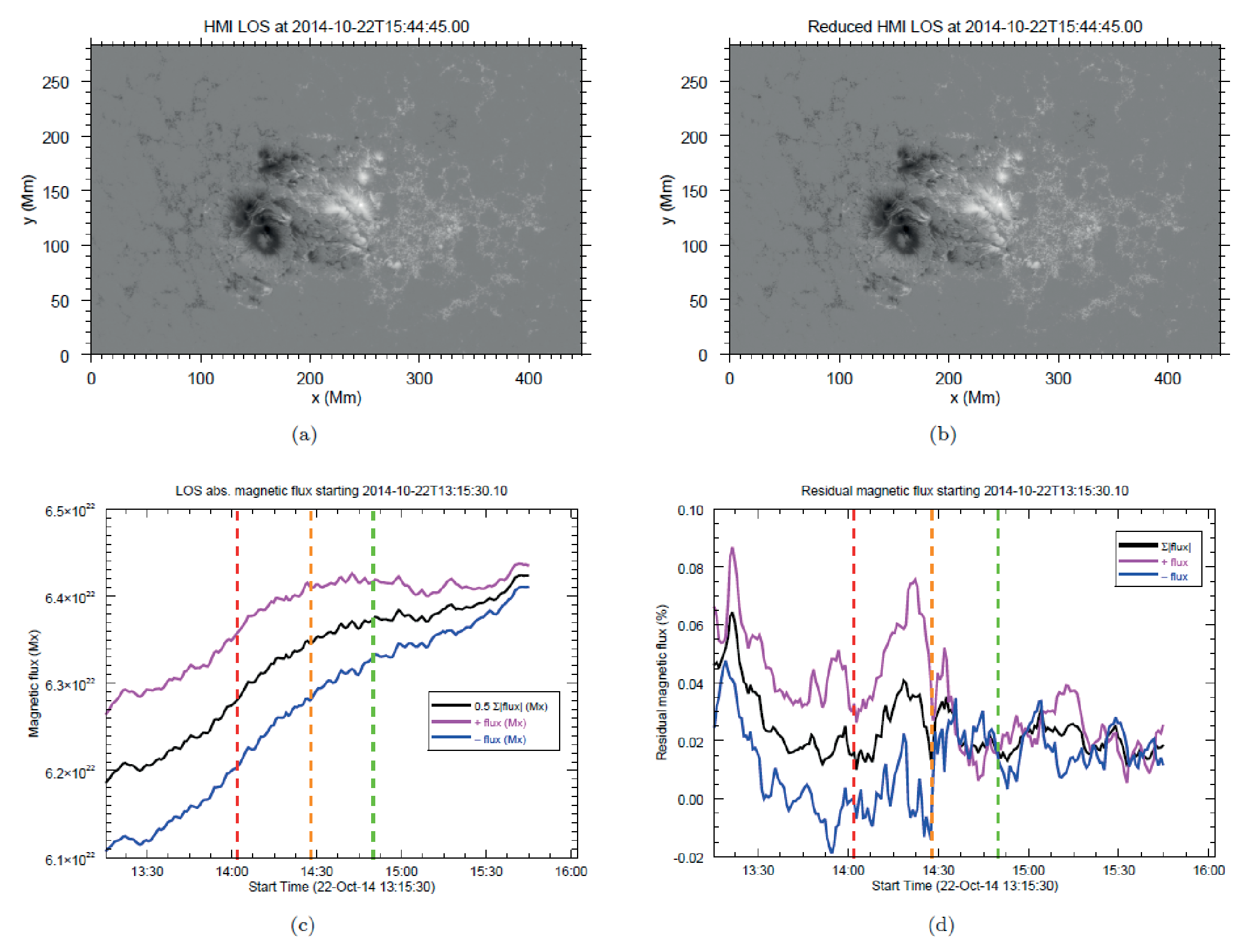}
\caption{Effect of spatial resolution for AR 12192 on 22 October 2014 for the X1.6 flare. (Top row) {\em SDO}/HMI magnetograms (a) at full resolution and (b) half resolution; (Bottom row) evolution of the total unsigned magnetic flux ($\times$0.5, black), positive flux (magenta) and negative flux (blue) for (c) the full resolution magnetogram, and (d) the residual flux between the two resolution magnetograms. The vertical lines indicates the flare timing (see Table~\ref{tab:events}, start (red), peak (orange), and end (green) times).}
\label{fig:ar12192_1}
\end{figure*} 

\section{Data description and methods}\label{sec:data}

	\subsection{Selected events}\label{sec:ar}

The datasets to analyse the correlations between X-class ﬂaring sites and the existence of MNPs are selected using the following criteria. We list the ﬂares of X class occurring between the years of 2010--2019 covering the majority of solar cycle 24. The Geostationary Operational Environmental Satellites (GOES) observations provide the soft X-ray flux in the range 1-8~\AA\ allowing to differentiate between the different ﬂare classes. We select all the events with a soft X-ray flux above 10$^{-4}$ W m$^{-2}$ corresponding to flare class above X1.0. We found a total of 51 events. These events indeed cover the entire solar cycle 24 (Dec. 2008 to Dec. 2019) as there are no X-class flare reported in 2008 and 2009 prior to the SDO mission. We also note that there are no X-class flare reported in 2010, 2016, 2018, and 2019.    

We restrict the number of events to the flares that are located in Active Regions (ARs) that are far enough from the limb to avoid projection effects in finding the location of flare's sites and most importantly for the line-of-sight magnetic field needed for the magnetic field extrapolations. Typically the selected AR is within $\sim$500\arcsec\ of disk centre. We also removed from the list ARs that possess large scale connectivity to other nearby ARs. The number of events satisfying those criteria is 17 X-class flares covering the entire solar cycle 24. In Table~\ref{tab:events}, the properties of the X-class flares are listed: date, timing of the flare (start, peak and end times), flare class (and soft X-ray flux), source active region, and the location of the active region at the time of the flare. For reference, we add the event number as listed by \citet{2017ApJ...834...56T} as the authors provided a comprehensive analysis of the photospheric magnetegrams, soft X-ray flux, flare ribbon properties and associated CMEs (see their Table 3) for our selected events. Our selection criteria have excluded all the X-class flares in 2015 and 2017. In 2015, there are two X-class flares: AR 12297 on March 11 too close to the limb, AR 12335 on May 5 within the required field-of-view but part of a larger group of three ARs. In 2017, there are 4 X-class flares on September 6--8 in AR 12673 which is located too close to the limb.

Selected X-class flares have been well studied from different perspectives, from multi-wavelength observations to MagnetoHydroDynamics (MHD) simulations: AR11158 \citep[e. g.][]{2011ApJ...738..167S,2012ApJ...745L..17W,2012ApJ...761..105L,2012ApJ...745L...4L,2012ApJ...749...85G,2012ApJ...759...50P,2013SoPh..287..415P,2013ApJ...770....4T,2013ApJ...772..115T,2015ApJ...803...73I}, 
AR11166 \citep[e. g.][]{2014ApJ...792...40V}, 
AR11283 \citep[e. g.][]{2015A&A...582A..55R,2018A&A...619A..65M,2020ApJ...903..129P,2021ApJ...916...37Z}, 
AR11429 \citep[e. g.][]{2015ApJ...809...34C,2016ApJ...817...14P,2020ApJ...901...40D}, 
AR11520 \citep[e. g.][]{2014ApJ...789...93C}, 
AR11890 \citep[e. g.][]{2016ApJ...820L..21X,2021ApJ...909..105T}, 
AR12017 \citep[e. g.][]{2014ApJ...796...85J,2016ApJ...824..148Y,2018ApJ...860..163W,2019ApJ...883...32C}, 
AR12158 \citep[e. g.][]{2020SoPh..295...87G,2021FrASS...8...35K,2023NatAs.tmp...94G}, 
AR12192 \citep[e. g.][]{2015ApJ...808L..24C,2016ApJ...818..168I,2017ApJ...840..116B,2017ApJ...845...54Z,2018ApJ...869..172L}, 
AR12242 \citep[e. g.][]{2021MNRAS.501.4703J}. Amongst the ARs listed above, other flaring events have been studied in great details. For instance, \citet{2019ApJ...878...78C} have studied the properties of the M8.7 flare on 2014 December 17 in AR 12242. The authors have reported that the magnetic configuration exhibited a magnetic null point associated with the reconnection process leading to quasi-periodic pulsations.

	\subsection{EUV data}\label{sec:aia}

The period of time 2010--2020 has been defined to correspond to the availability of high-resolution, high-cadence EUV observations provided by the Atmospheric Imaging Assembly (AIA, \citeauthor{2012SoPh..275...17L} \citeyear{2012SoPh..275...17L}) on board of the Solar Dynamics Observatory (SDO, \citeauthor{2012SoPh..275....3P} \citeyear{2012SoPh..275....3P}).  We use the 171~\AA\ passband centered on the Fe {\sc ix} emission line to locate the flaring sites by identifying the brightenings in the images and movies (often it is observed as a saturation of the CCD camera). The spatial resolution and time cadence are 1\farcs5 and 12 s, respectively. We produce movies in the 171~\AA, 193~\AA, 211~\AA, and 304~\AA\ passbands to analyse the time evolution of the events at different temperature ranges and to confirm the position of the flare sites. The location of the flare is taken as the first appearance of a strong brightening in the EUV images at 171~\AA\ and not the subsequent brightenings that are associated with post-flare loops. 

	\subsection{Magnetic field extrapolation}\label{sec:extrapol}

We investigate the structure of the magnetic polarities in the photosphere. We use the line-of-sight (LOS) magnetic field measurements provided by the Helioseismic and Magnetic Imager (HMI, \citeauthor{2012SoPh..275..207S} \citeyear{2012SoPh..275..207S}) on board SDO. Based on the timing of the events (see Table~\ref{tab:events}), we select a magnetogram of the ARs before the flare and one after the flare within a time window of 1-2 hours around the flare peak time. We select the {\em SDO}/HMI magnetograms which are not influenced by the flaring process, especially by removing the data that show instrumental effects due to the high-energy particles. For each magnetogram, we derive the 3D coronal magnetic field configurations using the potential field assumption. The magnetic field satisfies the following equations:
\begin{equation}
\nabla \times \vec{B} = \vec{0}, \quad \textrm{and} \quad
\nabla \cdot \vec{B} = 0.
\end{equation}
The extrapolations are done in Cartesian coordinates, where $x$ is the East-West direction, $y$ is the South-North direction, and $z$ is the vertical/radial direction \citep{1999A&A...350.1051A}. In the solar case, the extrapolation method for a potential field requires as boundary conditions the vertical/LOS magnetic field component in the photosphere corresponding to the $z = 0$ layer. This corresponds to the {\em SDO}/HMI magnetograms. The other sides of the computational box have open boundary conditions with only the normal component to the surface that does not vanish.  The {\em SDO}/HMI magnetograms have a pixel size of 0\farcs5 and a field-of-view that depends on the active region properties (between 175$\times$100 Mm and 450$\times$280 Mm). In order to optimize the computational time, we reduce the size of the magnetograms by a factor of 2 in both directions, limiting to computational boxes' size below 500 pixel$^3$. To illustrate the effect of modifying the pixel size, we investigate the changes in the total unsigned magnetic flux and the positive/negative magnetic fluxes for AR 12192 on 22 October 2014 (X1.6 flare). In Figs.~\ref{fig:ar12192_1}a and b, we plot the full-resolution and reduced-resolution magnetograms recorded on 22 October 2014 at 15:44:45 UT, after the X1.6  flare. The visual inspection of both magnetograms does not reveal any clear changes in the photospheric distribution of the magnetic polarities. The $x$ and $y$ axes are in the coordinate system of the computational box and expressed in Mm. In Fig.~\ref{fig:ar12192_1}c, we plot the evolution of the magnetic fluxes around and during the flare period: the total unsigned magnetic flux (black curve), the positive and negative magnetic fluxes (magenta and blue curves respectively). Fig.~\ref{fig:ar12192_1}c represents the fluxes for the full-resolution magnetogram with the vertical lines indicating the flare timing (see Table~\ref{tab:events}). The total unsigned flux varies from 6.2~10$^{22}$ Mx to 6.4~10$^{22}$ Mx during a period of $\sim$ 2 hours. In Fig.~\ref{fig:ar12192_1}d, we compute the residual fluxes between the two resolutions (in \%): the residual flux is always low during the period of time ($<$ 0.08\%). This analysis reveals that reducing the resolution of the magnetograms does not affect the variation of the magnetic flux and the spatial distribution of the polarities. Therefore the extrapolations at lower resolution will produce very similar/identical 3D magnetic field configurations.

	\subsection{Magnetic null point properties}\label{sec:mnp}

Once we have obtained the 3D magnetic field configurations from the potential field extrapolations, we investigate the existence and properties of Magnetic Null Points (MNPs). The finding of MNPs relies on the trilinear interpolation method developed by \citet{2007PhPl...14h2107H}. The advantage of the trilinear method is to be able to find MNPs in weakly nonlinear magnetic fields within grid cells as confirmed by the comparative study of different algorithms in \citet{2020A&A...644A.150O}. We used the method described by \citet{2012SoPh..277..131R} in order to classify MNPs based on their properties.

Based on the assumption that the magnetic ﬁeld, $\vec{B}$, around the MNP linearly approaches zero, we may express $\vec{B}$ as a ﬁrst-order Taylor expansion:
$$\vec{B} = M \vec{r}, $$
where $M$ is the Jacobian matrix comprising elements $\displaystyle M_{ij} = \frac{\partial B_j}{\partial x_i}$ for all $i, j = 1, 2, 3$ describing the spatial components, and $\vec{r}$ is the position vector $(x, y, z)$ in Cartesian coordinates. If $M$ is invertible, $M$ has three (real and/or complex) eigenvalues $(\lambda_1, \lambda_2, \lambda_3)$. Since the magnetic ﬁeld is constrained by the solenoidal condition, $\nabla \cdot \vec{B} = 0$, the trace of the Jacobian matrix satisfies $Tr(M) = 0$:
\begin{equation}
\lambda_1 + \lambda_2 + \lambda_3 = 0.
\label{eq:lambda}
\end{equation}

Eq.~\ref{eq:lambda} implies that there are either three real eigenvalues or one eigenvalue and two complex conjugate eigenvalues. We consider that $\lambda_1$ is the largest real eigenvalue, and $\lambda_2$ and $\lambda_3$ are two real or two complex conjugate eigenvalues. From Eq.~\ref{eq:lambda}, it is clear that the eigenvalues will satisfy $|\lambda_1| = |\lambda_2 + \lambda_3|$, which further implies that $|\lambda_1| > |\lambda_2|, |\lambda_3|$. These eigenvalues describe the direction of the spine ﬁeld line ($\lambda_1$) and fan plane ($\lambda_2$ and $\lambda_3$) associated to the MNP \citep{1997GApFD..84..245P}. The direction of the spine (and thus of the fan field lines) determines the MNP type: if the spine ﬁeld line is directed towards the MNP and the fan field lines spread out from the MNP, then the MNP is known as a type B or positive; similarly a type A or negative MNP consists of the ﬁeld lines pointing towards the MNP in the fan plane and directed away from the MNP along the spine. The skeleton structure of the MNP is then deﬁned by the vectors associated with the eigenvalues.

\begin{figure*}
\includegraphics[width=0.45\linewidth]{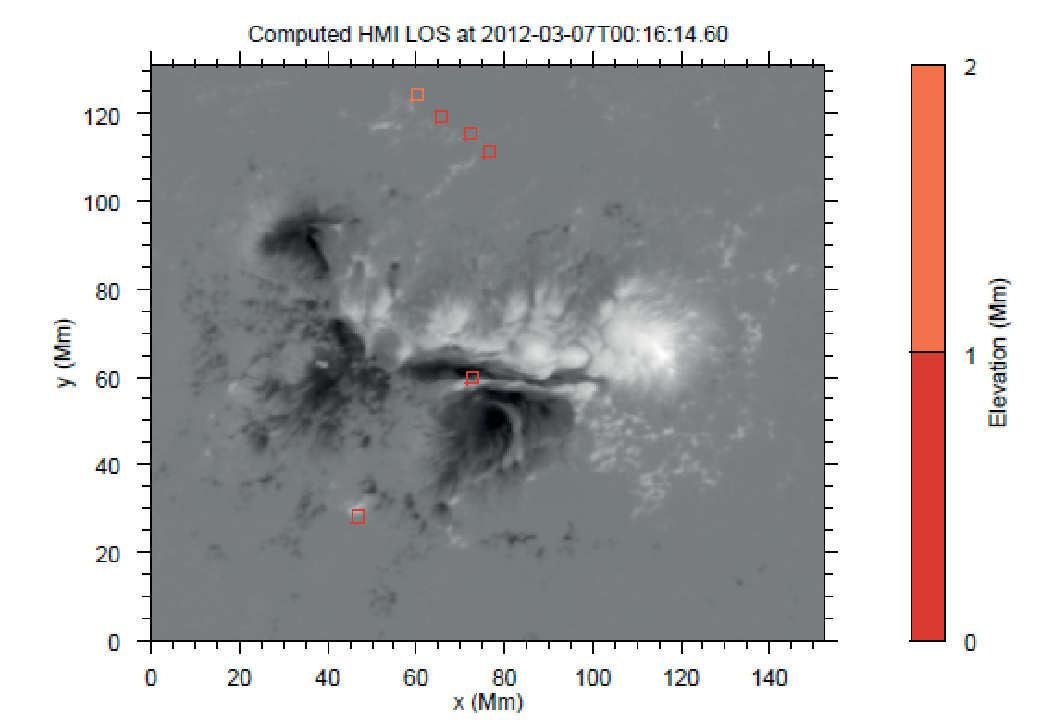}
\includegraphics[width=0.485\linewidth]{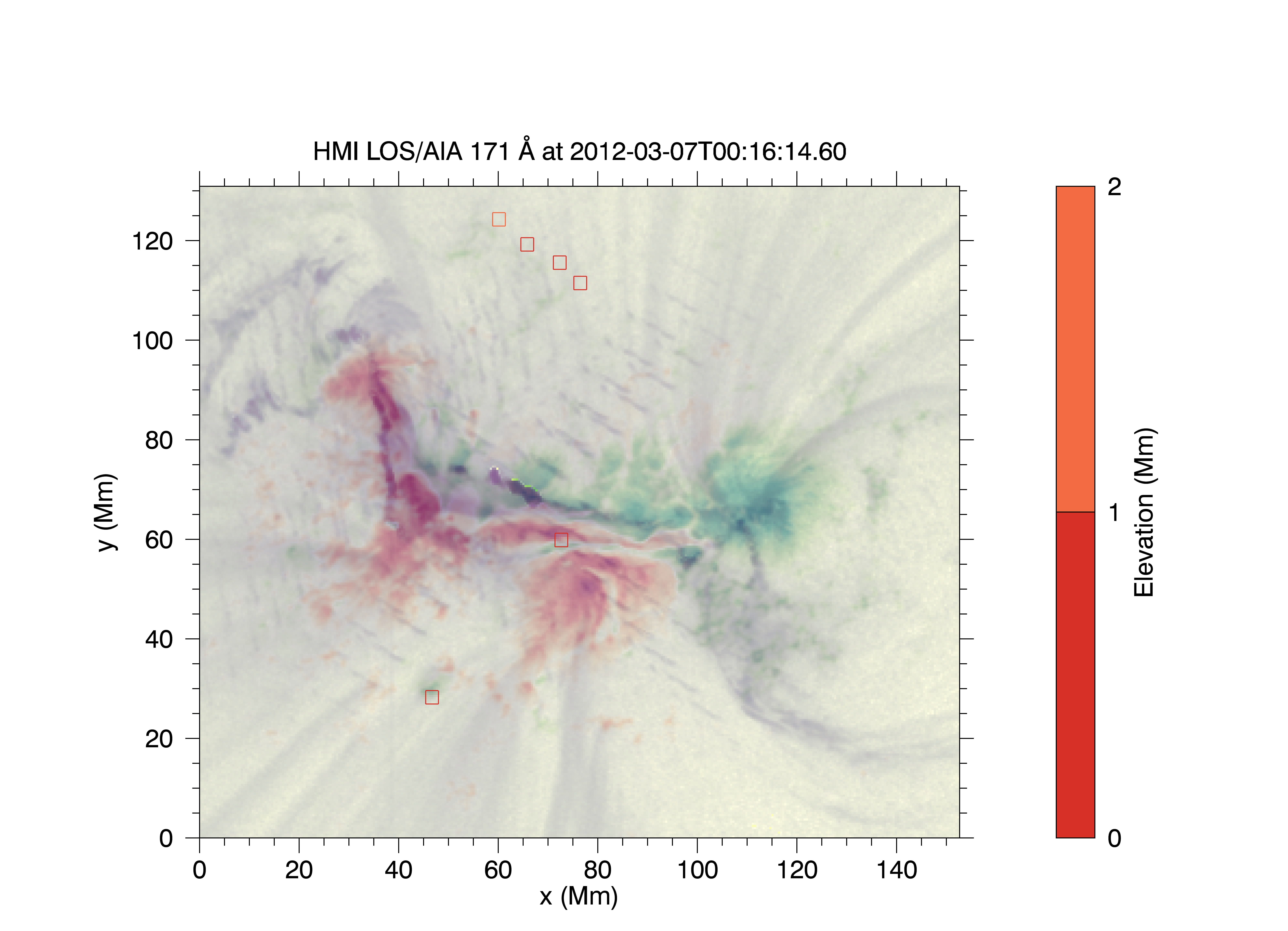}
\caption{MNPs detected in active region AR 11429 on 7 March 2012: (left) {\em SDO}/HMI photospheric magnetic field at 00:16 UT, (right) overlaid of the {\em SDO}/HMI magnetogram (blue and red colours for positive and negative LOS magnetic field respectively) with the {\em SDO}/AIA image at 171\AA\ (purple colours) at the same time. The MNPs are represented by coloured squares corresponding to their elevation/height above the photospheric magnetogram (colour bar).}
\label{fig:ar11429_1}
\end{figure*}

Complementing the eigen properties of the Jacobian matrix, \citet{2012SoPh..277..131R} introduced the spectral radius, $\rho_J$, of the Jacobian matrix $M$:
\begin{equation}
\rho_J = \max{(|\lambda_i|)} \quad \textrm{for} \quad i=1,2,3
\label{eq:rhoj}
\end{equation}
where $\lambda_i$ are the eigenvalues of $M$. The spectral radius may be regarded as a good proxy for the stability of a null point in a magnetic-ﬁeld conﬁguration \citep{2012SoPh..277..131R}. In addition, (relatively) large spectral radii indicate large magnetic ﬁeld gradients in the vicinity of the MNP. Where the eigenvalues of the Jacobian matrix are complex, the spectral radius is the real eigenvalue, satisfying the divergence-free property of the magnetic ﬁeld. It is important to note that complex eigenvalues correspond to non-physical MNPs that cannot exist under a force-free assumption (including potential field), but may be numerically obtained due to a failure of the algorithm used to locate them or due to the lost of validity of the linear assumption (and hence the representation of the magnetic ﬁeld as a first order Taylor expansion) around the MNP \citep{2012SoPh..277..131R}.

For a potential magnetic field, the eigenvalues are real and the fan plane is orthogonal to the spine field line. The spine direction is provided by the eigenvectors for the largest eigenvalue and the fan plane is defined by the other two eigenvectors. We also introduce the angle $\theta_{spine}$ between the spine field line and the vertical direction $z$ (altitude) in the Cartesian potential field extrapolations. 

\citet{2012SoPh..277..131R} has demonstrated that MNPs with the largest spectral radius are stable in magnetic configurations subject to different force-free assumptions (from potential to non-linear force-free) and for different amount and sign of current densities. The stability of MNPs is also used to differentiate between true MNPs that are stable and artificial MNPs that are unstable and associated with the noise of the photospheric magnetogram. Consequently, it is expected that a MNP with large spectral radius will not be affected by the slow evolution of the magnetic configuration and/or by a reconnection process.

\section{Results}\label{sec:res}

    \subsection{Properties of MNPs}
We determine the location of all MNPs for the selected events (see Table~\ref{tab:events}) before and after the peak of the soft X-ray flux in the 1-8\AA\ wavelength range: only the MNP with the largest spectral radius is considered for the association.  We correlate the location of the MNPs to the flare site identified in EUV images (mostly in the 171~\AA\ channel of {\em SDO}/AIA as seen in Fig.~\ref{fig:ar11429_1} right).

In Table~\ref{tab:res}, we list all the MNPs associated with X-class flares, at or near to flare site. For each computation before and after the event, we specify (third column) if a MNP exists near the flare site considering that the MNP is within a circle of radius 10 pixels around the flare site. If a MNP exists, we provide the height (in Mm), the type (positive or negative), the three eigenvalues, the properties of the spine field line and fan plane, the angle $\theta_{spine}$ between the spine field line and the vertical direction, a schematic of the spine field line (in red) in the ($x$, $y$, $z$) frame of reference, and the value of the photospheric LOS magnetic field component just below the MNP. We distinguish two categories of events: those associated with a MNP (or more than one MNP) at or near the flare site, and those not associated with a MNP. For the MNPs classified as not associated with the flaring site, we did check the {\em SDO}/AIA movies in the different wavelength bands to ascertain that there is no association with the MNPs within the extrapolated fields during  the flaring process. 

In Table~\ref{tab:nonassos}, we list the remaining events that are not associated with a MNP near the flaring site. Again only the MNPs with the largest spectral radius are considered.  

	\subsection{Example of AR11429}


In Fig.~\ref{fig:ar11429_1}, we take the example of the X5.3 flare observed in AR 11429 on 7 March 2012. The location of the MNPs is represented by the squares and the colour indicates the height of the MNPs above the photospheric layer. There are 6 MNPs detected in the 3D potential magnetic field configuration obtained from the magnetogram (background image). From Fig.~\ref{fig:ar11429_1} (left), the 4 MNPs located in the Northern part of the image are within a quiet region outside the core of the active region. The other two MNPs are contained in the core of the active region. Fig.~\ref{fig:ar11429_1} (right) confirms that the MNP located at (75, 60) Mm is associated with the flare site as the origin of the diffraction pattern observed in the EUV channel at 171~\AA, and is above a region of strong magnetic field strength of about 600 G (see Table~\ref{tab:res}), i.e. above a polarity inversion line (PIL). This MNP has the largest spectral radius. The magnetic configuration which contains four polarities (2 positive and 2 negative) and three PILs in the East-West direction is often associated with a reversed Y-shape configuration that contains a MNP and near vertical spine field line (see also \citeauthor{2005ESASP.596E..61R} \citeyear{2005ESASP.596E..61R}). This fact is confirmed by the direction of the eigenvector in Table~\ref{tab:res}. Its height is 0.54$\pm$0.04 Mm. 

	\subsection{Flares associated with MNPs}\label{sec:withmnp}

From Table~\ref{tab:res}, 6 flare sites are related to at least one MNP: X1.5 on 9 March 2011, X5.4 on 7 March 2012, X1.3 on 7 March 2012, X3.3 on 5 November 2013, X1.1 on 10 November 2013, and X1.0 on 29 March 2014. The X1.1 flare on 10 November 2013 occurring in AR 11890 has three MNPs in its vicinity and thus we consider their correlation with the flare site in the statistical analysis. Only 35\% of the X-class flare sites are directly associated with MNPs. In two events, the MNP disappears after the flare. In one case, the MNP appears after the flare. The MNPs are located relatively close to the photospheric surface: all MNPs are below 2 Mm but one which is located at 4.4 Mm. The MNP associated with a flare site has the largest spectral radius $\rho_{J}$ of all MNPs found within the magnetic field configurations. It indicates that the flares occur in places where magnetic gradients in the neighbourhood of MNP are the strongest, and thus where large electric currents are located. 

The before and after locations of the MNPs are largely unchanged, except for the negative MNP. The combination of those results regarding the sign of the MNP seems to indicate that the magnetic reconnection process might differ for positive or negative MNPs in the corona. This is counter-intuitive as there is no difference in theory between a positive and a negative MNP.   

\begin{table*}
	\centering
	\caption{Characterisation of X-class flares associated with a magnetic null point}
	\label{tab:res}
	\begin{tabular}{ccccccp{2.9cm}ccc} 
		\hline
		\hline
		Date & Timing & MNP & Height & Type & Eigenvalues & \centering Spine and Fan  & $\theta_{spine}$ & Geometry & $B_{phot}$ \\
		& & & (Mm) & & ($\lambda_1 = \pm \rho_J$, $\lambda_2$, $\lambda_3$) & \centering Eigenvectors ($x, y, z$) & with $z$ & & (G) \\
		\hline
		 2011 Mar 9 & Before & Y & 1.69 & + & (-304.2, 43.1, 266.3) & \centering (0.086,0.176, 0.981) \newline (0.890, 0.435, 0.135) \newline (-0.312, 0.814, -0.485)& 11$\fdg$4 & \begin{minipage}{.07\textwidth}
		 	\centering
			\begin{tikzpicture}[scale=0.6]
  			\draw[->, gray] (0,0,0) -- (0,0,1) node[left] {$x$};
 		 	\draw[->, gray] (0,0,0) -- (1,0,0) node[above] {$y$};
  			\draw[->, gray] (0,0,0) -- (0,1,0) node[above] {$z$};
			\draw[<-, red, line width=0.4mm](0,0,0) -- (0.176, 0.981, 0.086);
			\end{tikzpicture} 
		 \end{minipage} & +405 \\[-0.3cm]
		 & After & N & & & & & & \\
		 2012 Mar 7 & Before & Y & 0.57 & + & (-847.7, 89.2, 599.3) & \centering (0.041, 0.170, 0.985) \newline \centering (0.967, 0.219, -0.130) \newline \centering (-0.218, 0.940, -0.261) &  10$\fdg$2 & \begin{minipage}{.07\textwidth}
		 	\centering
			\begin{tikzpicture}[scale=0.6]
  			\draw[->, gray] (0,0,0) -- (0,0,1) node[left] {$x$};
 		 	\draw[->, gray] (0,0,0) -- (1,0,0) node[above] {$y$};
  			\draw[->, gray] (0,0,0) -- (0,1,0) node[above] {$z$};
			\draw[<-, red, line width=0.4mm](0,0,0) -- (0.170, 0.985, 0.041);
			\end{tikzpicture} 
		 \end{minipage} & +587 \\[-0.5cm]
		  & After & Y & 0.50 & + & (-789.3, 86.0, 535.1) & \centering (0.091, 0.213, 0.973) \newline \centering (0.962, 0.067, -0.264) \newline \centering (-0.123, 0.921, -0.368) & 13$\fdg$4 & \begin{minipage}{.07\textwidth}
		 	\centering
			\begin{tikzpicture}[scale=0.6]
  			\draw[->, gray] (0,0,0) -- (0,0,1) node[left] {$x$};
 		 	\draw[->, gray] (0,0,0) -- (1,0,0) node[above] {$y$};
  			\draw[->, gray] (0,0,0) -- (0,1,0) node[above] {$z$};
			\draw[<-, red, line width=0.4mm](0,0,0) -- (0.213, 0.973, 0.091);
			\end{tikzpicture} 
		 \end{minipage} & +549 \\[-0.5cm]
		 2012 Mar 7 & Before & Y & 0.54 & + & (-835.1, 41.2, 634.3) & \centering (0.086, 0.176, 0.981) \newline \centering (0.967, 0.130, -0.218) \newline \centering (-0.163, 0.950, -0.266) & 11$\fdg$4 & \begin{minipage}{.07\textwidth}
		 	\centering
			\begin{tikzpicture}[scale=0.6]
  			\draw[->, gray] (0,0,0) -- (0,0,1) node[left] {$x$};
 		 	\draw[->, gray] (0,0,0) -- (1,0,0) node[above] {$y$};
  			\draw[->, gray] (0,0,0) -- (0,1,0) node[above] {$z$};
			\draw[<-, red, line width=0.4mm](0,0,0) -- (0.176, 0.981, 0.086);
			\end{tikzpicture} 
		 \end{minipage} & +615  \\[-0.5cm]
		  & After & Y & 0.50 & + & (-763.5, 69.8, 536.9) & \centering (0.093, 0.213, 0.972) \newline \centering  (0.964, 0.075, -0.255) \newline \centering (-0.137, 0.921, -0.365) & 13$\fdg$3 &  \begin{minipage}{.07\textwidth}
		 	\centering
			\begin{tikzpicture}[scale=0.6]
  			\draw[->, gray] (0,0,0) -- (0,0,1) node[left] {$x$};
 		 	\draw[->, gray] (0,0,0) -- (1,0,0) node[above] {$y$};
  			\draw[->, gray] (0,0,0) -- (0,1,0) node[above] {$z$};
			\draw[<-, red, line width=0.4mm](0,0,0) -- (0.213, 0.972, 0.093);
			\end{tikzpicture} 
		 \end{minipage} & +553 \\
		 2013 Nov 5 & Before & Y & 0.50 & $-$ & (177.6, -132.3, -61.0) & \centering (0.631, 0.163, 0.758) \newline \centering (0.513, 0.434, -0.740) \newline \centering (-0.485, 0.816, 0.315) & 40$\fdg$7 & \begin{minipage}{.07\textwidth}
		 	\centering
			\begin{tikzpicture}[scale=0.6]
  			\draw[->, gray] (0,0,0) -- (0,0,1) node[left] {$x$};
 		 	\draw[->, gray] (0,0,0) -- (1,0,0) node[above] {$y$};
  			\draw[->, gray] (0,0,0) -- (0,1,0) node[above] {$z$};
			\draw[->, red, line width=0.4mm](0,0,0) -- (0.163, 0.758, 0.631);
			\end{tikzpicture} 
		 \end{minipage} & $-$19 \\[-0.5cm]
		  & After & Y & 1.16 & $-$ & (177.3, -145.1, -41.4) & \centering (0.424, 0.308, 0.851) \newline \centering (0.758, 0.384, -0.528) \newline \centering (-0.509, 0.857, -0.077) & 31$\fdg$6 & \begin{minipage}{.07\textwidth}
		 	\centering
			\begin{tikzpicture}[scale=0.6]
  			\draw[->, gray] (0,0,0) -- (0,0,1) node[left] {$x$};
 		 	\draw[->, gray] (0,0,0) -- (1,0,0) node[above] {$y$};
  			\draw[->, gray] (0,0,0) -- (0,1,0) node[above] {$z$};
			\draw[->, red, line width=0.4mm](0,0,0) -- (0.308, 0.851, 0.424);
			\end{tikzpicture} 
		 \end{minipage} & $-$206 \\
		 2013 Nov 10 & Before & Y & 1.93 & + & (-40.3, 6.5, 34.0) & \centering (0.070, -0.543, 0.837) \newline \centering (0.807, 0.521, 0.279) \newline \centering (-0.567, 0.660, 0.493) & 33$\fdg$2 & \begin{minipage}{.07\textwidth}
		 	\centering
			\begin{tikzpicture}[scale=0.6]
  			\draw[->, gray] (0,0,0) -- (0,0,1) node[left] {$x$};
 		 	\draw[->, gray] (0,0,0) -- (1,0,0) node[above] {$y$};
  			\draw[->, gray] (0,0,0) -- (0,1,0) node[above] {$z$};
			\draw[<-, red, line width=0.4mm](0,0,0) -- (-0.543, 0.837, 0.070);
			\end{tikzpicture} 
		 \end{minipage} & +22 \\[-0.5cm]
		  & After & Y & 1.90 & + & (-46.2, 14.5, 33.5) & \centering (-0.016, -0.497, 0.868) \newline \centering (0.749, 0.551, 0.367) \newline \centering (-0.650, 0.663, 0.371) & 29$\fdg$8 & \begin{minipage}{.07\textwidth}
		 	\centering
			\begin{tikzpicture}[scale=0.6]
  			\draw[->, gray] (0,0,0) -- (0,0,1) node[left] {$x$};
 		 	\draw[->, gray] (0,0,0) -- (1,0,0) node[above] {$y$};
  			\draw[->, gray] (0,0,0) -- (0,1,0) node[above] {$z$};
			\draw[<-, red, line width=0.4mm](0,0,0) -- (-0.497, 0.868, -0.016);
			\end{tikzpicture} 
		 \end{minipage} & +52 \\[-0.5cm]
		 & Before & Y & 4.41 & + & (-23.2, 9.0, 14.1) & \centering (0.032, 0.711, -0.703) \newline \centering (0.964, -0.222, -0.147) \newline \centering (0.288, 0.662, 0.692) & -45$\fdg$4 & \begin{minipage}{.07\textwidth}
		 	\centering
			\begin{tikzpicture}[scale=0.6]
  			\draw[->, gray] (0,0,0) -- (0,0,1) node[left] {$x$};
 		 	\draw[->, gray] (0,0,0) -- (1,0,0) node[above] {$y$};
  			\draw[->, gray] (0,0,0) -- (0,1,0) node[above] {$z$};
			\draw[<-, red, line width=0.4mm](0,0,0) -- (0.711, -0.703, 0.032);
			\end{tikzpicture} 
		 \end{minipage} & +5 \\[-0.5cm]
		  & After & Y & 4.40 & + & (-23.2, 8.9, 14.1) & \centering (0.048, 0.706, -0.707) \newline \centering (0.963, -0.234, -0.134) \newline \centering (0.285, 0.664, 0.691) & -45$\fdg$0 & \begin{minipage}{.07\textwidth}
		 	\centering
			\begin{tikzpicture}[scale=0.6]
  			\draw[->, gray] (0,0,0) -- (0,0,1) node[left] {$x$};
 		 	\draw[->, gray] (0,0,0) -- (1,0,0) node[above] {$y$};
  			\draw[->, gray] (0,0,0) -- (0,1,0) node[above] {$z$};
			\draw[<-, red, line width=0.4mm](0,0,0) -- (0.706, -0.707, 0.048);
			\end{tikzpicture} 
		 \end{minipage} & +5 \\
		  & Before & N & & & & & & & \\[-0.5cm]
		  & After & Y & 0.81 & + & (-56.2, 21.4, 24.4) & \centering (0.288, -0.698, 0.656) \newline \centering (0.905, 0.099, -0.424) \newline \centering (0.206, 0.702, 0.680) &  49$\fdg$0 & \begin{minipage}{.07\textwidth}
		 	\centering
			\begin{tikzpicture}[scale=0.6]
  			\draw[->, gray] (0,0,0) -- (0,0,1) node[left] {$x$};
 		 	\draw[->, gray] (0,0,0) -- (1,0,0) node[above] {$y$};
  			\draw[->, gray] (0,0,0) -- (0,1,0) node[above] {$z$};
			\draw[<-, red, line width=0.4mm](0,0,0) -- (-0.698, 0.656, 0.288);
			\end{tikzpicture} 
		 \end{minipage} & +19 \\[-0.5cm]
		 2014 Mar 29 & Before & Y & 0.67 & + & (-348.1, 82.1, 230.8) & \centering (-0.287, 0.377, 0.881) \newline \centering (0.860, -0.03, 0.509) \newline \centering (0.168, 0.821, -0.545) & 37$\fdg$2 & \begin{minipage}{.07\textwidth}
		 	\centering
			\begin{tikzpicture}[scale=0.6]
  			\draw[->, gray] (0,0,0) -- (0,0,1) node[left] {$x$};
 		 	\draw[->, gray] (0,0,0) -- (1,0,0) node[above] {$y$};
  			\draw[->, gray] (0,0,0) -- (0,1,0) node[above] {$z$};
			\draw[<-, red, line width=0.4mm](0,0,0) -- (0.377, 0.881, -0.287);
			\end{tikzpicture} 
		 \end{minipage} & +154 \\
		  & After & N & & & & & & & \\
		\hline
	\end{tabular}
\end{table*}

From the results in Table~\ref{tab:res}, we notice that a vast majority of MNPs associated with a flare site is of positive sign and with a spine field line almost vertical/radial: 87.5\% of the MNPs are positive. All three eigenvalues are also listed to show that the divergence free assumption is correct to a good accuracy: the deviation from the divergence-free magnetic field is from the interpolation method used and from the degree of non-linearity in the vicinity of the MNP. The vertical component of the photospheric magnetic field corresponding to the orthogonal projection of the MNP onto the photosphere is of the same sign as the MNP.

	\subsection{Non-association}\label{sec:nonmnp}

\begin{table}
	\centering
	\caption{Properties of null points not associated with X-class flares}
	\label{tab:nonassos}
	\begin{tabular}{cccc} 
		\hline
		\hline
		Date & Timing & Height & Type \\
		& & (Mm) &  \\
		\hline
		2011 Feb 15 & Before & 1.18 & $-$ \\
		& After & 8.39 & $-$ \\
		2011 Sep 6 &  Before & 0.56 & + \\
		& After & 0.46 & + \\
		2011 Sep 7 &  Before & 0.48 & $-$\\
		& After & 1.20 & + \\
		2012 Jul 12 &  Before & 0.51 & + \\
		& After & 0.46 & $-$ \\
		2013 Nov 8 &  Before & 1.98 & + \\
		& After & 2.40 & + \\
		2014 Sep 10 &  Before & 1.55 & $-$ \\
		& After & 1.08 & $-$ \\
		2014 Oct 22 &  Before & 1.00 & $-$ \\
		& After & 1.00 & $-$ \\
		2014 Oct 24 &  Before & 2.73 & $-$ \\
		& After & 3.00 & $-$ \\
		2014 Oct 25 &  Before & 1.83 & $-$ \\
		& After & 1.73 & $-$ \\
		2014 Oct 26 &  Before & 0.96 & $-$ \\
		& After & 1.44 & $-$ \\
		2014 Dec 20 &  Before & 1.20 & + \\
		& After & 1.58 & + \\
		\hline
	\end{tabular}
\end{table}

In Table~\ref{tab:nonassos}, we list the remaining 11 X-class flares within 7 different active regions as well as the properties of the MNPs having the largest spectral radius. There is no spatial correlation between the MNP location and the flaring region as observed in the {\em SDO}/AIA EUV channels. Those MNPs are located in the outskirt of the active region far from the flaring site. We note that a large number of MNPs are of negative type: 55\% are negative, 27\% are positive, and 18\% are changing sign before and after the flare.

In this sample of active regions, there are statistically more X-class flares that are not associated with a MNP. 


\section{Discussion and conclusions}\label{sec:concl}

By combining photospheric magnetic field data, EUV coronal data, and potential field magnetic extrapolations, we have shown that during solar cycle 24, 35\% of the selected X-class flares are associated with a magnetic null point (MNP). Of these events, 87.5\% MNPs are of positive type, i.e. spine field line oriented towards the MNP, and the spine field line is mostly in the vertical/radial direction. This implies that the fan field lines are radiating away from the MNP. All the MNPs associated with the X-class flares have the largest spectral radius, indicating the presence of strong magnetic field gradients and/or large electric currents.      

The existence of a spine field line that is almost vertical has been reported in previous studies of large flares (M-class and X-class) using a combination of magnetic field extrapolations and multi-wavelength observations \citep[][e.g.]{2013ApJ...778..139S}. This particular magnetic topology has been shown to be significant for triggering a magnetic reconnection process \citep{2022LRSP...19....1P}. Regarding the sign of the MNP, in theory the physics of MNPs does not depend on the sign (positive or negative). Based on this consideration, the studies of 3D magnetic reconnection found in the literature are largely done for an analytical solution of a positive MNP \citep[see e.g.][]{2009PhPl...16l2101P}.

Understanding the magnetic topology and configurations involved in large flares is a step forward to contribute the development of Space Weather prediction, moving from photospheric magnetograms to 3D magnetic field extrapolations. It is also primordial to understanding the underlying physics processes of eruptions. Our findings are unfortunately not conclusive to narrow down the possible trigger mechanisms and thus have a good proxy for Space Weather. The striking results that a large percentage of MNPs associated with flares are positive MNPs is a step forward to predict flares that are related to MNPs, however, it does not constrain enough any of the flare mechanisms: the existence of a MNP with a large spectral radius is a clear sign of magnetic energy storage in the corona, but other storage/release processes exist such as in a complex quasi-topology (without MNP) and/or with a twisted flux bundle. As it is clear from \citet{2017ApJ...834...56T}, the events associated with a MNP are not always associated with a CME and/or a geomagnetic storm. The existence of a MNP in a flaring active region does not prevail in anyway on the eruption mechanism or the consequences of the flare (e.g. CME, confined flare, filament eruption).


\section*{Acknowledgments}
The authors would like to thank the Royal Astronomical Society for providing support to this project through the student bursary scheme. Data used in this article are courtesy of NASA/SDO and the AIA and HMI science teams. Data have been downloaded from the Virtual Solar Observatory (VSO). 

\section*{Data Availability}
The HMI and AIA data underlying this article are publicly available in any {\em SDO} data centre, or through the Virtual Solar Observatory at https://sdac.virtualsolar.org/cgi/search. The potential field extrapolation datasets will be shared on reasonable request to the corresponding author.



\bibliographystyle{mnras}
\bibliography{topo_bib}








\bsp	
\label{lastpage}
\end{document}